\documentclass[twocolumn,english,amssymb,prl]{revtex4-1}
\usepackage[T1]{fontenc}
\usepackage[latin9]{inputenc}
\usepackage{color}
\usepackage{amsmath}
\usepackage{amssymb}
\usepackage{graphicx}
\PassOptionsToPackage{normalem}{ulem}
\usepackage{ulem}

\makeatletter

\providecommand{\tabularnewline}{\\}

\@ifundefined{textcolor}{}
{%
 \definecolor{BLACK}{gray}{0}
 \definecolor{WHITE}{gray}{1}
 \definecolor{RED}{rgb}{1,0,0}
 \definecolor{GREEN}{rgb}{0,1,0}
 \definecolor{BLUE}{rgb}{0,0,1}
 \definecolor{CYAN}{cmyk}{1,0,0,0}
 \definecolor{MAGENTA}{cmyk}{0,1,0,0}
 \definecolor{YELLOW}{cmyk}{0,0,1,0}
 }

\graphicspath{{FiguresMain/}}
\@ifundefined{definecolor}
 {\usepackage{color}}{}
\usepackage{dcolumn}% Align table columns on decimal point
\usepackage{bm}% bold math
\usepackage{caption}
\newcounter{one}
\makeatother

\usepackage{babel}
\begin{document}

\title{Multipartite Einstein-Podolsky-Rosen
steering and genuine tripartite entanglement with optical networks}

\author{Seiji Armstrong$^{1*}$, Meng Wang$^{2}$, Run Yan Teh$^{3}$, Qihuang Gong$^{2}$,
Qiongyi He$^{2,3*}$, Jiri Janousek$^{1}$, Hans-Albert Bachor$^{1}$, Margaret D. Reid$^{3*}$,
Ping Koy Lam$^{1,4}$}

\affiliation{$^{1}$Centre for Quantum Computation and Communication Technology,
Department of Quantum Science, The Australian National University,
Canberra, ACT 0200, Australia\\
 $^{2}$State Key Laboratory for Mesoscopic Physics, School of
Physics, Peking University, and Collaborative Innovation Center of Quantum Matter, Beijing, China\\
 $^{3}$Centre for Quantum and Optical Science, Swinburne University of Technology,
Melbourne, Victoria 3122, Australia\\
$^{4}$College of Precision Instrument and Opto-Electronics Engineering, Tianjin University, Key Laboratory of Opto-Electronics Information Technology, Ministry of Education, Tianjin, 300072, China.\\
*email: seiji.armstrong$@$gmail.com; qiongyihe$@$pku.edu.cn; mdreid$@$swin.edu.au.}

\maketitle
%\begin{abstract}
%This is my abstract
%\end{abstract}

\textbf{The Einstein-Podolsky-Rosen (EPR) paradox \cite{epr} established
a link between entanglement \cite{steering-1,ent} and nonlocality in
quantum mechanics \cite{nonlocality}. EPR steering \cite{eprsteereric,Wiseman,wise2,steering}
is the nonlocality associated with the EPR paradox and has traditionally
only been investigated between two parties \cite{ou epr,hann group,smithsteerxp,bwzeil,hw-1,rrmp-1,boyd}.
Here, we present the first experimental observations of multipartite
EPR steering, and of the genuine tripartite continuous variable entanglement of three mesoscopic optical systems \cite{threeent,aokicv, eiscv}. We explore different linear optics networks - each
one with optimised asymmetries - that create multipartite steerable
states containing different numbers of quantised optical modes (qumodes). By introducing asymmetric loss on a 7-qumode state, we characterize 8 regimes of directional steering, showing that $N+1$ regimes exist for an $N$-qumode state. Further, we reveal the directional monogamy of steering, and experimentally demonstrate continuous variable one-sided semi device-independent quantum secret sharing \cite{secretsh}. Our methods establish principles for the development of multiparty quantum communication protocols with asymmetric observers, and can be extended to qubits, whether photonic \cite{svetexp,bwzeil,hw-1,boyd,threeent,smithsteerxp}, atomic \cite{14blatt}, superconducting \cite{supercondqubits}, or otherwise.}\\

Schr{\"o}dinger introduced the term {}``steering\textquotedblright{}
to describe the nonlocality apparent in the EPR paradox, and pointed
out that these states involve a quantum property called {}``entanglement''
\cite{steering,steering-1}. Wiseman et al \cite{Wiseman,wise2} have
formalised the meaning of steering in terms of violations of local
hidden state models, and revealed that the EPR paradox is a manifestation
of quantum steering. In simple terms, quantum steering dictates that
measurements made by one observer can apparently {}``steer'' (alter) the
state of another observer at a different location. 

The observation of multipartite EPR steering has not been possible
until recently as the framework necessary to understand the concept
has only just been developed  \cite{eprsteereric,Wiseman,wise2,genuineEPR}. 
The motivation to expand this framework arises from considerations of real-world quantum networks, such as the quantum internet \cite{KimbleNet}, for which security and privacy are of paramount importance. Here, we expand on the theoretical framework and derive optimised criteria to detect multipartite EPR steering using linear optical circuits. The criteria involve the
canonical position and momentum observables, which are realised in our experiment as
highly efficient quadrature phase amplitude measurements.
Following the criteria, we present the first experimental investigation
of multipartite EPR steering, including demonstration of directional
\emph{monogamy} relations which give bounds on the way steering is distributed
among the different parties. Further, we demonstrate the principle of one-sided device-independent
quantum secret sharing and in doing so confirm for the first time
the continuous variable genuine tripartite entanglement of three optical
modes. For bipartite EPR states, there are 3 different regimes: 2-way,
1-way, and no-way steering \cite{onewaysteer,steermurray}. In
general, for each $N$ qumode state, $N+1$ regimes of
steering are possible. Here, we create 7 different quantum networks, each $ $producing a multipartite
EPR steerable state, with different levels of correlations. By introducing
asymmetry into the network, we manipulate the 7-qumode state, to
experimentally achieve all 8 different regimes of directional steering.

Underpinning the idea of multipartite EPR steering is the quantum concept of entanglement \cite{ent,steering-1}. $N$ systems are
genuinely $N$-partite entangled if and only if the entanglement of
the $N$-party system cannot be produced by mixing quantum states with fewer than $N$ systems entangled \cite{threeent, eiscv}. Suppose three observers (Alice,
Bob and Charlie) each make measurements on three respective quantum
systems, labelled $1,2,$ and $3$. We show in the Supplementary Materials
that genuine tripartite entanglement of the three systems is confirmed
if:
\begin{eqnarray}
\Delta(\hat{x}_{1}-\frac{(\hat{x}_{2}+\hat{x}_{3})}{\sqrt{2}})\,\,\,\,\,\,\,\,\,\,\,\,\,\,\,\,\,\,\,\,\,\,\,\,\,\,\,\,\nonumber \\
\times\Delta(\hat{p}_{1}+\frac{(\hat{p}_{2}+\hat{p}_{3})}{\sqrt{2}})<1.\label{eq:genentxp}
\end{eqnarray}
Here $\hat{x}_{i}$,$\hat{p}_{i}$ ($i=1,2,3$) are the position and
momentum observables of the system $i$, scaled in such a way that
the Heisenberg uncertainty relation becomes $\Delta\hat{x_{i}}\Delta\hat{p_{i}}\geq1$. Genuine tripartite entanglement has a different meaning to full tripartite inseparability \cite{threeent, eiscv}. The latter occurs when the entanglement cannot be produced by entangling any (fixed) two parties, and does not eliminate that entanglement is created by mixing different bipartite entangled states. 

To understand tripartite EPR steering, we again consider that Alice's measurements
are the observables $\hat{x}_{1}$ and $\hat{p}_{1}$ of a quantum
system. However, this assumption is no longer applied to Bob and Charlie, who need not report the results of  quantum observables. We suppose that Bob
and Charlie can collaborate to give a prediction for the
outcome of Alice's $\hat{x}_{1}$ (or $\hat{p}_{1}$) measurement
and denote the average uncertainty in their inferences by $\Delta_{inf}\hat{x}_{1}$
(and $\Delta_{inf}\hat{p}_{1}$). If $\Delta_{inf}\hat{x}_{1}\Delta_{inf}\hat{p}_{1}<1$,
then we realise an EPR steering paradox \cite{mreidepr,Wiseman}.
In that case, it is as though Alice's $\hat{x}_{1}$ and $\hat{p}_{1}$
values were predetermined to an accuracy that contradicts quantum
mechanics \cite{epr}, or else that there is an actual {}``steering''
of Alice's system by Bob and Charlie's actions \cite{steering}. We
symbolise this directional EPR steering by the notation $BC\rightarrow A$. Confirming steering of A by the group BC amounts to confirming entanglement between the two groups, but with fewer assumptions made about group BC. Steering is therefore a greater experimental challenge than entanglement.

Next, we extend to $N$ observers and consider genuine tripartite steering. Any $N-$party state demonstrates
$N$-partite EPR steering of the $j$th site by the remaining set of
sites denoted $K$ if 
\begin{equation}
S_{j|K}\equiv\Delta_{\mathrm{inf}}(\hat{x}_{j})\Delta_{\mathrm{inf}}(\hat{p}_{j})<1.\label{eq:EPR}
\end{equation}

We define the the square of this product, $(S_{j|K})^{2}$, to be the EPR steering number. In our experiment, the steering measurements are optimised linear
combinations of the  $\hat{x}_{k}$ ($\hat{p}_{k}$) ($k\neq j$). Thus

\begin{eqnarray*}
\Delta_{\mathrm{inf}}(\hat{x}_{j}) & = & \Delta(\hat{x}_{j}+\sum_{k\neq j}g_{k,x}x_{k}),\\
\Delta_{\mathrm{inf}}(\hat{p}_{j}) & = & \Delta(\hat{p}_{j}+\sum_{k\neq j}g_{k,p}p_{k}),
\end{eqnarray*}
where $g_{x,k}$ and $g_{p,k}$ are optimised real numbers. The steering detected by equation (\ref{eq:EPR}) is genuinely $N$-partite
if it cannot be explained as arising from any steering limited to
$N-1$ or fewer parties. In the tripartite case, let $A,B,C$ be the
sites of Alice, Bob and Charlie. All fixed two-party steering is negated,
if we can demonstrate each of \cite{genuineEPR} 
\begin{equation}
S_{A|BC}<1,S_{B|AC}<1,S_{C|AB}<1,\label{eq:genuinetri}
\end{equation}
which implies steering across all bipartitions: $BC\rightarrow A$, $AC\rightarrow B$, and $AB\rightarrow C$. The condition
(\ref{eq:genuinetri}) confirms the full inseparability of the quantum
density matrix and also of any three-party hidden state model that
could describe the system \cite{Wiseman}. The
negation does not however rule out that the steering has been created
by mixing states with two-party steering across different bipartitions.
To claim genuine tripartite steering, we need to eliminate this possibility
\cite{genuineEPR}. We prove in the Suppplementary Materials that
this is done if
\begin{eqnarray}
\Delta(\hat{x}_{1}-\frac{(\hat{x}_{2}+\hat{x}_{3})}{\sqrt{2}})\,\,\,\,\,\,\,\,\,\,\,\,\,\,\,\,\,\,\,\,\,\,\,\,\,\,\,\,\nonumber \\
\times\Delta(\hat{p}_{1}+\frac{(\hat{p}_{2}+\hat{p}_{3})}{\sqrt{2}})<0.5\label{eq:gensteer}
\end{eqnarray}
which is a stricter form of inequality Eq.~\eqref{eq:genentxp}.

\begin{figure*}[ht]
\centering{}\includegraphics[clip,width=15cm]{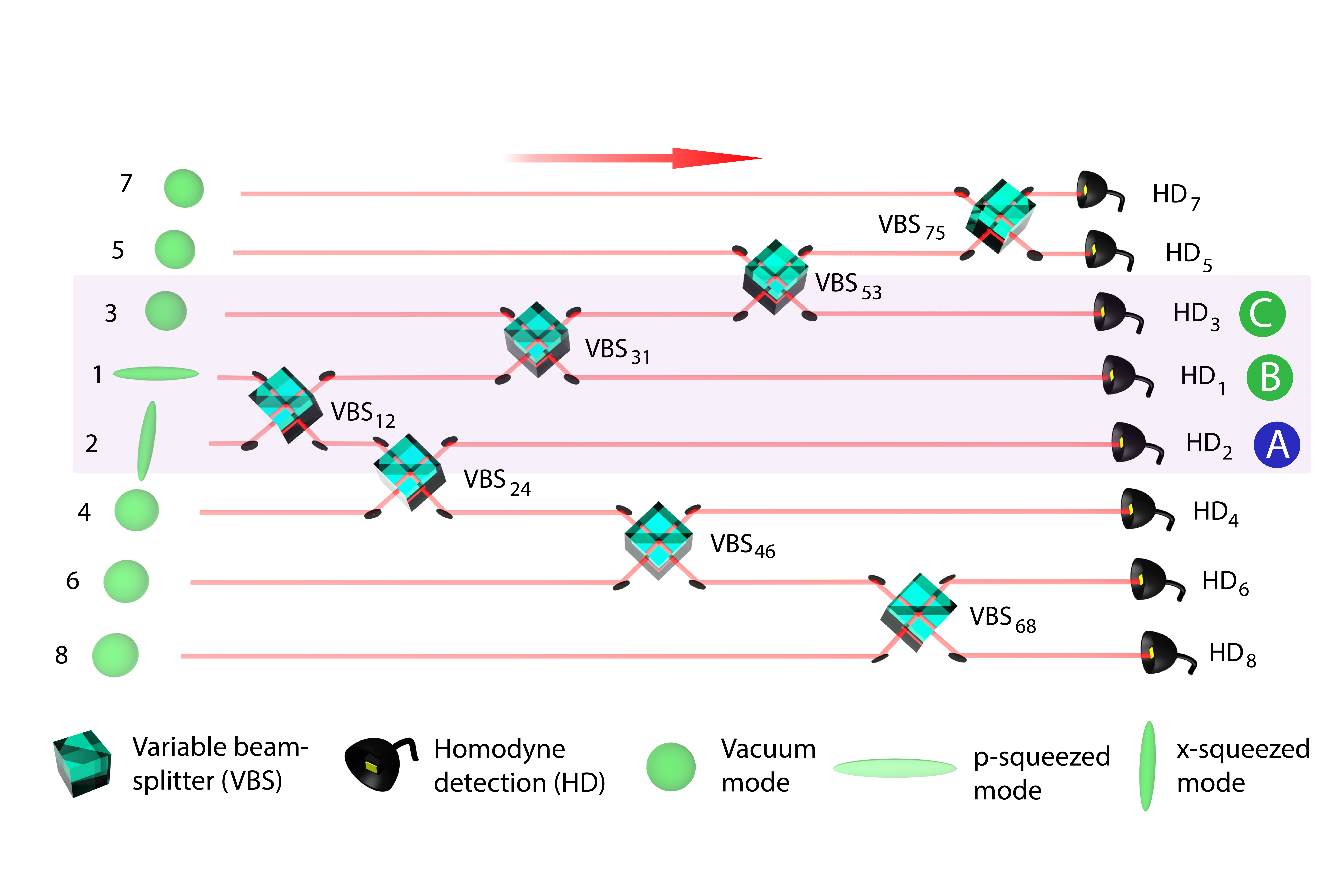} \caption{\textbf{Programmable linear optics circuit for demonstrating multipartite steering.}
The circuit employs 2 quadrature-squeezed qumodes and 6 quantum-noise limited vacuum modes as inputs. Each of the 7 beam-splitters are fully tuneable from $R=0\%$ to $R=100\%$. Beam-splitters are optimised for inputs with asymmetric squeezing.
\label{FigSteeringCircuit} }
\end{figure*}

We investigate multipartite entanglement and steering by employing the programmable linear optics circuit developed in ref~\cite{Armstrong2012} in order to create various multi-partite quantum states from different networks.
Independent qumodes are shaped in order to be multiplexed on the same beam. By programmatically changing the measurement basis, the scheme allows us to emulate linear optics networks in real time. The various quantum networks that we create for this demonstration can be visualised in the programmable circuit of Fig.~\ref{FigSteeringCircuit}.
We input 2 quadrature-squeezed qumodes and 6 quantum-noise limited vacuum modes into the linear optics circuit, and we have the freedom to programmatically vary each beam-splitter's reflectivity. We set a beam-splitter reflectivity to $R=100\%$ and set it to a perfect mirror when we choose not to mix a particular input qumode into the state. 

\begin{figure*}[ht]
\centering{}\includegraphics[clip,width=15cm]{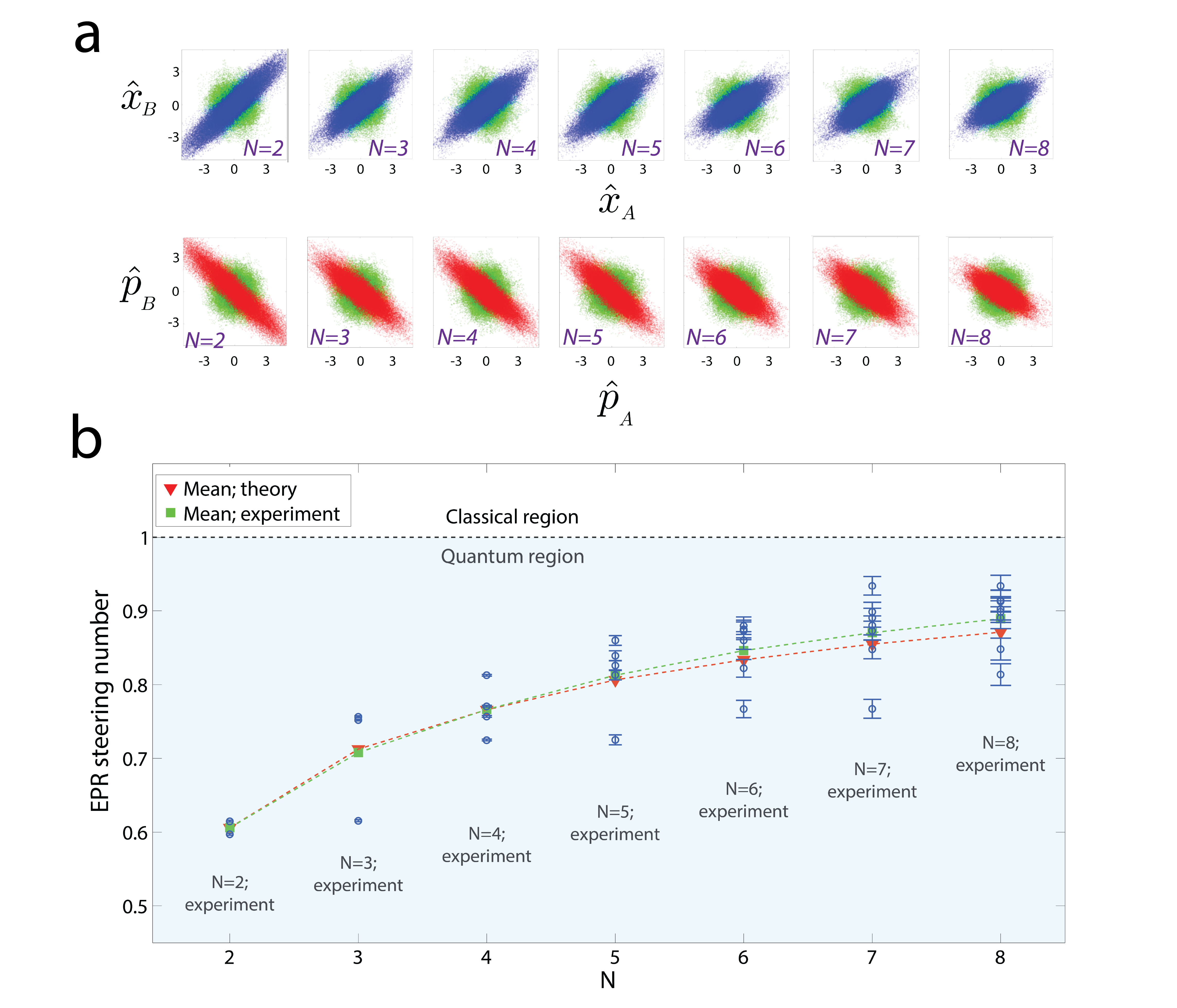} \caption{\textbf{Multipartite EPR Steering for the $N$-partite entangled states,
$N=2-8$.} Here ``Classical'' means that this condition cannot be obtained for any optimised choice of gains $g_{k,x}$ and $g_{k,p}$.
(a) Normalised quadrature amplitudes in $\hat{x}$ ($\hat{p}$) are plotted parametrically in order to visualise correlations quantified by equation~\eqref{eq:EPR} of main text. For all 7 states, the $\hat{x}$ ($\hat{p}$) quadrature amplitude of the first qumode in the $N$-qumode state is plotted against the linear combination of the $\hat{x}$ ($\hat{p}$) quadrature amplitudes of the remaining qumodes. Green data points show shot noise measurements, blue (red) data points show $\hat{x}$ ($\hat{p}$) quadrature measurements. Correlations weaken as $N$ increases.
(b) The blue markers give for each $N$-partite state the $N$
experimental values for the EPR steering number $\left(S_{j|K}\right)^{2}$,
where we consider the steering of the $j$th qumode ($j=1,..,N$) by
the remaining group $K$. The condition $\left(S_{j|K}\right)^{2}<1$
of inequality (\ref{eq:EPR}) for EPR steering of qumode $j$ is satisfied
in each case. The experimental values are consistent with the theoretical predictions (red markers) explained in the Methods Section.
\label{FigParametric} }
\end{figure*}

Homodyne detection is employed to measure the inferred variances of each qumode in the state. This provides highly efficient detection, and our measurements of the amplitudes do not rely on fair sampling assumptions \cite{smithsteerxp,bwzeil}. Each qumode is characterised by the conjugate quadrature operators $\hat{x}$ and $\hat{p}$ of the quantum harmonic oscillator mode of the light field. Optimal circuits are created that take into account asymmetries in squeezing values of the input qumodes. 

In our experiment we input two mixed states with different magnitudes of squeezing, allowing us to explore how best to bias and optimise the beam-splitter networks,

\begin{align}
\begin{pmatrix}(\Delta\hat{p}_{1})^{2}\\
(\Delta\hat{x}_{1})^{2}\\
(\Delta\hat{p}_{2})^{2}\\
(\Delta\hat{x}_{2})^{2}
\end{pmatrix}=\begin{pmatrix}-3.6\pm0.05dB\\
+8.9\pm0.05dB\\
-4.1\pm0.05dB\\
+9.5\pm0.05dB
\end{pmatrix}
\end{align}

The optimised linear optics networks that we create for our mixed inputs are detailed in the methods section. We first generate 7 different quantum states from 7 different networks, with each beam-splitter's reflectivity given in Table~\ref{table:BS1}.

\begin{table}[h]
\caption{Beamsplitter reflectivities for optimised optics networks.}
\begin{tabular*}{0.5\textwidth}{@{\extracolsep{\fill}}@{\extracolsep{\fill}}@{\extracolsep{\fill}}lcccccccr}
\textbf{N}  & \textbf{VBS$_{12}$}  & \textbf{VBS$_{31}$}  & \textbf{VBS$_{24}$}  & \textbf{VBS$_{53}$}  & \textbf{VBS$_{46}$}  & \textbf{VBS$_{75}$}  & \textbf{VBS$_{68}$}  & \tabularnewline
\hline 
\textbf{2}  & 50  & 100  & 100  & 100  & 100  & 100  & 100  & \tabularnewline
\textbf{3}  & 51.1  & 50  & 100  & 100  & 100  & 100  & 100  & \tabularnewline
\textbf{4}  & 50  & 50  & 50  & 100  & 100  & 100  & 100  & \tabularnewline
\textbf{5}  & 50.8  & 33.3  & 50  & 50  & 100  & 100  & 100  & \tabularnewline
\textbf{6}  & 50  & 33.3  & 33.3  & 50  & 50  & 100  & 100  & \tabularnewline
\textbf{7}  & 50.6  & 25  & 33.3  & 33.3  & 50  & 50  & 100  & \tabularnewline
\textbf{8}  & 50  & 25  & 25  & 33.3  & 33.3  & 50  & 50  & \tabularnewline
\hline 
\end{tabular*}
\label{table:BS1} 
\end{table}

The EPR correlations in each state can be visualised by parametrically
plotting components of the inferred variance terms. In all plots of
Fig.~\ref{FigParametric}, the $x$ axis is the $\hat{x}$ ($\hat{p}$)
quadrature amplitude of the first qumode in the state, and the $y$
axis is the linear combination of the $\hat{x}$ ($\hat{p}$) quadrature
amplitudes of the remaining $N-1$ qumodes in the $N$-qumode state.

\begin{figure*}
\centering{}\includegraphics[clip,width=15cm]{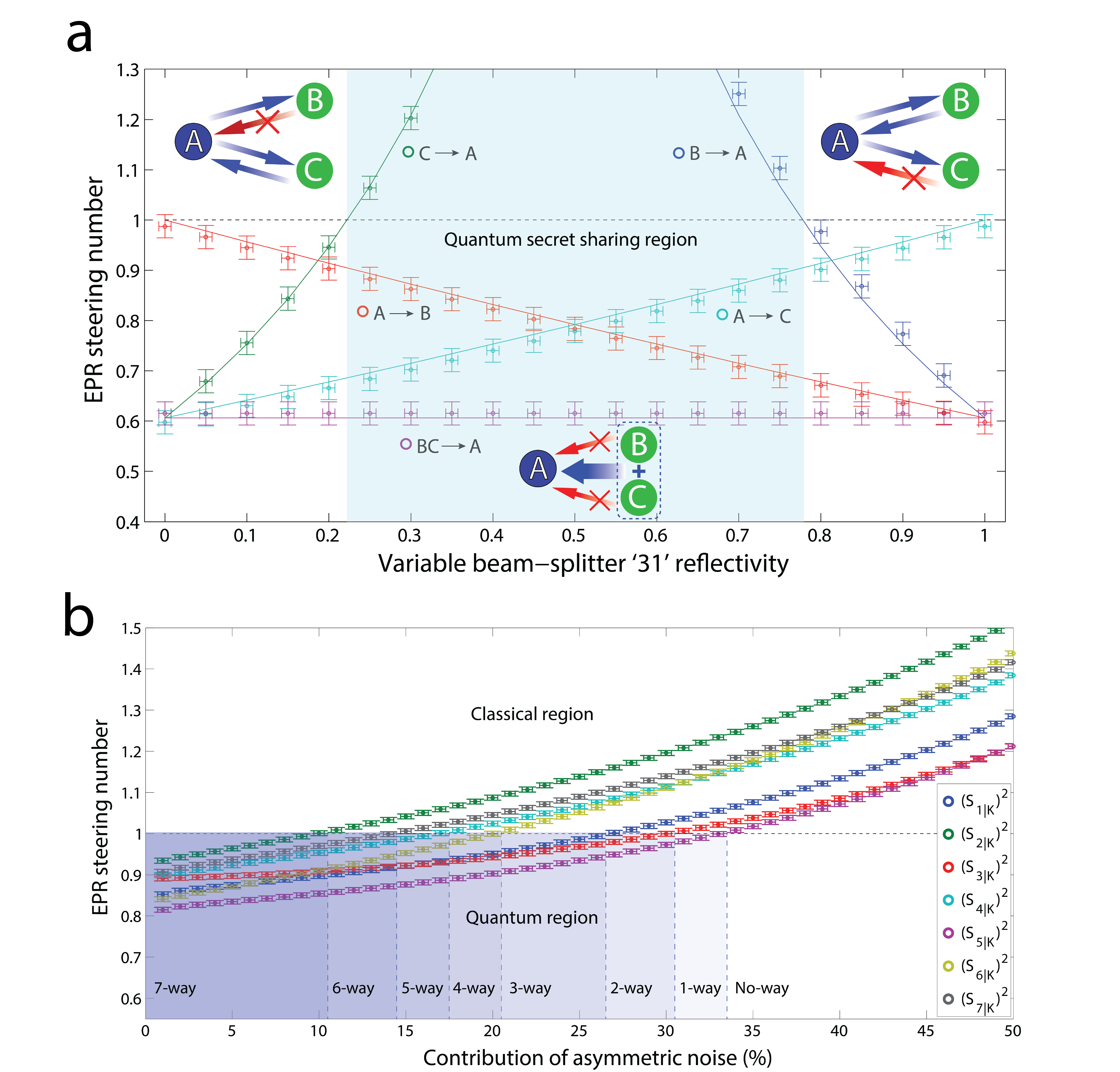} \caption{\textbf{Directional properties of multipartite steerable states.}
(a) EPR steering numbers for $N=3$ versus the reflectivity $R_{31}$
of the second beam splitter $VBS_{31}$. First beam-splitter $VBS_{12}$ held constant at $R_{12}=0.511$, and $R_{24}=R_{53}=1$. Markers indicate experimental data, curves are theoretical predictions. The single qumode on one side of the first beam splitter is denoted by $A$; the two qumodes on the other side by $B$ and $C$. Directional monogamy, where two parties ($B$ and $C$) cannot simultaneously independently steer the same third system ($A$), is shown by the blue and red arrows in the cartoons. Thus, steering of $A$ by $B$ (or $C$) is only possible for asymmetric beam splitters (blue and green markers). We confirm $S_{A|B}S_{A|C}\geq1$. However, the simultaneous steering of $B$ and $C$ by $A$ occurs for all $R_{31}$ (red and cyan markers). One-sided device-independent quantum secret sharing is shown in the blue regime of more symmetric beam splitters, where neither Bob nor Charlie can independently steer $A$ (red arrows), but they can steer $A$ by collaboration (magenta markers). (b) Manipulating the directional steering
of a $7$-partite steerable state by introducing asymmetric noise. 8 different regimes are demonstrated. Details are given in the Supplementary Materials.}
\label{fig:secret-sharing}
\end{figure*}

Perfect correlations would correspond to a semi-major axis of infinite
length, strictly on the diagonal. This would require infinite energy
and is unphysical; the ellipticity of each ellipse in Fig.~\ref{FigParametric} is indicative
of finite squeezing. We see that the higher the number of qumodes
in the state, the weaker the correlations become. This is evident
in the slight rotation off the strictly diagonal axis in both quadratures,
as well as the diminishing ellipticity of the correlations.
This is due to the additional vacuum contributions in our circuit
as we go to higher mode numbers.

Figure~\ref{FigParametric}b quantifies the correlations in
terms of the EPR steering number, Eq.~\eqref{eq:EPR}. For each $N$-qumode state,
there are $N$ EPR steering numbers that must be tested in order to confirm
the multipartite EPR steering of each qumode. Each EPR steering number represents
the steering from one partition of the state to the other, or the
direction of steering within the state. For $N=3$,
we confirm steering across all bipartitions, to satisfy criterion
(\ref{eq:genuinetri}), thus ruling out a large class of separable
classical models.

\begin{figure*}
\centering{}\includegraphics[width=12cm]{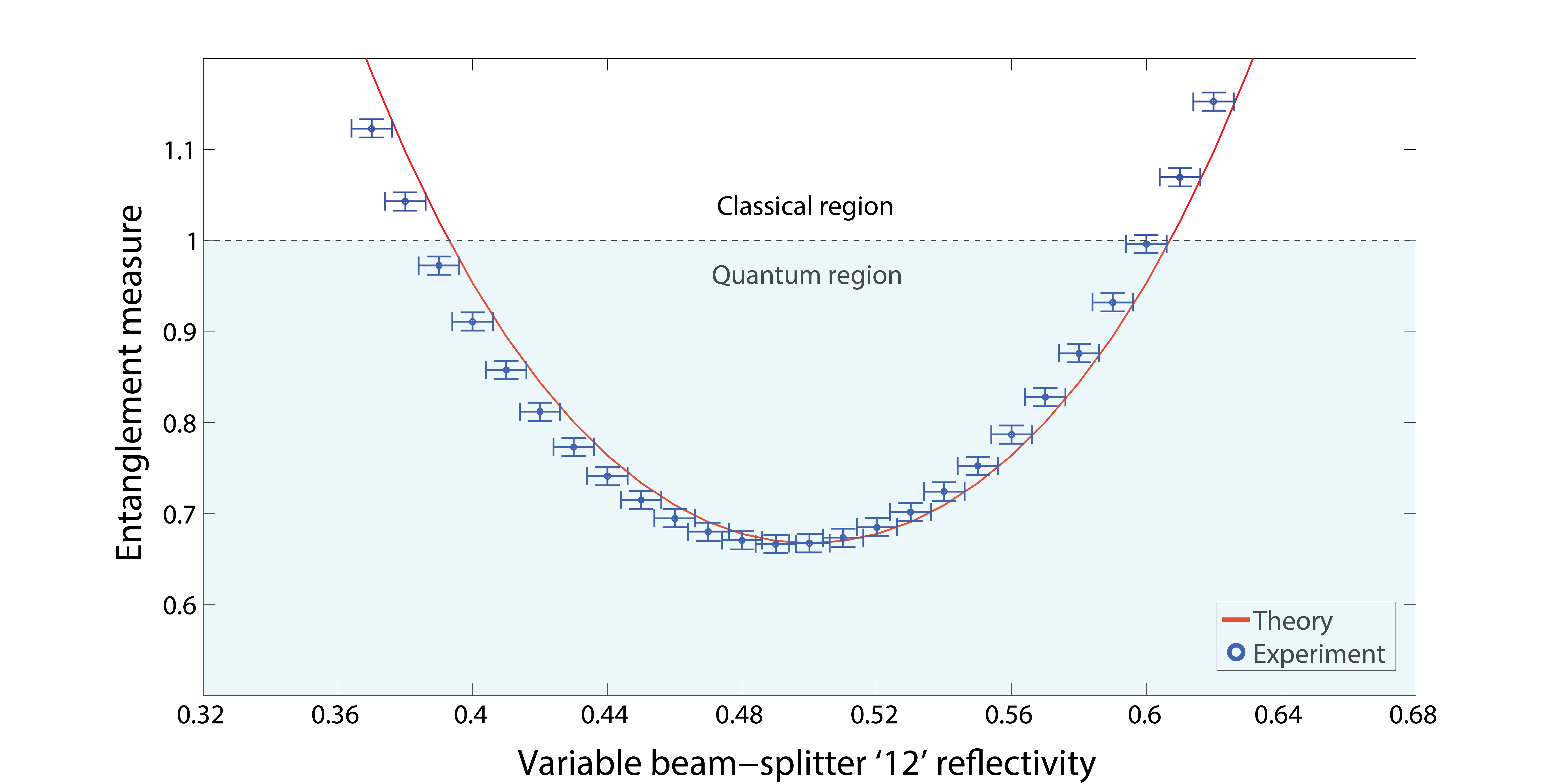} \caption{\textbf{Genuine tripartite entanglement. }The values of the genuine
tripartite entanglement parameter $\Delta^2(\hat{x}_{1}-\frac{(\hat{x}_{2}+\hat{x}_{3})}{\sqrt{2}})\Delta^2(\hat{p}_{1}+\frac{(\hat{p}_{2}+\hat{p}_{3})}{\sqrt{2}})$ as given by equation~\eqref{eq:genentxp} for the tripartite steerable state studied in Fig.~\ref{fig:secret-sharing}. A balanced beam-splitter is employed for the second beam-splitter in the network. The experimental data (blue circles) shows agreement with the predictions of the theoretical model (red curve) outlined in the Methods Section. The data satisfies the inequality (\ref{eq:genentxp}) and therefore confirms the genuine tripartite entanglement of three optical modes. We note that for $R_{1}\neq0.5$, the genuine tripartite entanglement can be more sensitively detected using a criterion involving an asymmetric choice of amplitude weightings (see Supplementary Materials). The condition for genuine tripartite steering (equation (\ref{eq:gensteer})) is predicted once the magnitude of squeezing of the input states is increased (see Supplementary Materials).}
\label{fig:Genuine-tripartite-entanglement}
\end{figure*}

We next demonstrate directional properties of multipartite EPR steering, which have implications for security in quantum communications. Firstly, the manifestation of steering
by inequality (\ref{eq:EPR}) is monogamous (Fig.~\ref{fig:secret-sharing}a). This means that if such steering is
shared between two parties, Alice and Bob, it cannot be shared between
Alice and a third party, Charlie. Mathematically, the monogamy is
described by $S_{A|B}S_{A|C}\geq1$ \cite{genuineEPR}. The EPR steering
number $(S_{A|B})^{2}$ therefore measures the directional security
of the $A-B$ channel, since it gives a lower bound on the inferred
uncertainty $S_{A|C}$ of Alice's amplitudes for any eavesdropping
parties {}``Eve'' denoted by $C$. As no assumptions are made about
the measurements of the steering parties, the security is independent
of the exact nature of Bob or Eve's measurement devices (though Alice's
measurement station must be trusted) \cite{onesidedcry}. Another
property of multipartite EPR steering is that for some regimes, the
steering of $A$ cannot take place by Bob (or Charlie) alone, but
requires both parties. In Fig.~\ref{fig:secret-sharing}a we illustrate
one-sided device-independent quantum secret sharing \cite{secretsh}
$-$ the values of Alice's amplitudes can only be unlocked with a
low uncertainty if the steering parties collaborate. 

Continuing for higher $N$, we modified the 7-qumode network to allow
the systematic introduction of asymmetric noise into the network. This is done by adding variable loss to one half of the network via a vacuum coupling beam-splitter acting on one output arm of VBS$_{12}$. The monogamy relation explains the impossibility of steering $S_{A|B}<1$
when the losses on the steering channel are $50\%$ or greater \cite{bowen}. Steering is also sensitive to the noise on the steered system.
Thus, one can manipulate the asymmetries to successively
disable the steering of each one of the parties (Fig.~\ref{fig:secret-sharing}b).
In this way, we find that all 8 separate steering regimes are accessible
by introducing up to $33\%$ of asymmetric loss into the circuit. The imperfect mode matching and detection (over $98\%$ efficiencies) of each qumode together with the systematic loss allows for the separate regimes.

Finally, the beam-splitters were varied to switch
from multipartite steering to genuine tripartite entanglement. In
Fig.~\ref{fig:Genuine-tripartite-entanglement}, we confirm the
genuine tripartite entanglement of three qumodes. The difference between
full tripartite inseparability \cite{aokicv,Armstrong2012} and genuine
tripartite entanglement has been explained  \cite{threeent}.
While the latter has been recently realised for three spatially separated
photons \cite{threeent}, our results demonstrate genuine tripartite
entanglement in a very different scenario: namely, for the EPR observables
of three fields consisting of many photons detected at very high efficiencies.

We have presented the first experimental evidence of multipartite EPR steering in various quantum states containing different levels of distributed squeezing. For the three-qumode state, we have established both the full inseparability of the hidden state model and genuine tripartite entanglement. Our work reveals properties of multipartite steering that link the amount of steering to the security of channels in the network, and further shows how the steering along different channels can be controlled. The framework developed opens up the possibility of demonstrations of multipartite EPR steering in various quantum network applications.\\ 

\textbf{Methods}\\

The linear optics networks used to generate the results presented in Fig.~\ref{FigParametric}a,b are characterised by the beam-splitters tabulated in table~\ref{table:BS1}.

The familiar bipartite EPR state is generated by setting the first beam-splitter reflectivity to $R=50\%$ and all other beam-splitters to mirrors. The measurement returns an EPR state and 6 unmixed vacuum modes, which are discarded. In order to create an $N$-qumode state we set $N-1$ of these to function as beam-splitters and the remaining $8-N$ to function as mirrors. We emphasise that although our qumodes are in principle spatially distinguishable, measurement events are not spacelike separated, and our work cannot address locality loopholes \cite{bwzeil}.

In all of our networks, when the two squeezed qumodes are pure states and equal in squeezing magnitude, the optimal beam-splitter reflectivity of the first beam-splitter is $R=50\%$. While unsurprising for quantum states containing even qumode numbers, it is less intuitive for odd numbered states, as one might expect that an unbalanced beam-splitter will favour the unbalanced network. The asymmetry is balanced by the quadrature amplitude optimisation gains.

For two mixed states that are unequal in squeezing values, the optimal beam-splitter ratio for even numbered states remains $R=50\%$. The symmetry arises from the EPR steering criteria being directional. The symmetry breaks down for mixed-state inputs when we consider odd-numbered quantum states. In this situation we benefit from biasing the beam-splitters.

It is only necessary to optimise the first beam-splitter of the network in odd numbered quantum states. We see that the reflectivities are $51.1\%$, $50.8\%$, and $50.6\%$ for the $3-$qumode, $5-$qumode, and $7-$qumode states, respectively. For any asymmetry in the inputs this will converge to $50\%$ for large $N$. 

We calculate the moments corresponding to the criteria above, using
the unitary transformation $a_{out,1}=\sqrt{\eta}a_{in,1}+\sqrt{(1-\eta)}a_{in,2}$,
$a_{out,2}=-\sqrt{(1-\eta)}a_{in,1}+\sqrt{\eta}a_{in,2}$ to model
the interaction of the qumodes at the beam splitter \cite{aokicv}.
Here, $a_{out,1}$, $a_{out,2}$ are the two output qumodes and $a_{in,1}$,
$a_{in,2}$ are the two qumodes input to the beam splitter. The reflectivity
of the beam splitter is given by $R=1-\eta$. The optimisation gain parameters are tabulated in the Supplementary materials.\\

\textbf{Author contributions}

S.A., P.K.L, Q.Y.H. and M.D.R. conceived of and designed the experiment.
S.A. and J.J. constructed and performed the experiment with supervision from H.-A.B. and P.K.L. 
S.A., M.W., R.Y.T., Q.H.G., Q.Y.H., and M.D.R. contributed equally to the theory. 
S.A. designed and created the virtual networks, and conducted the data analysis. 
S.A., Q.Y.H., and M.D.R. wrote the manuscript, and all authors commented on drafts.

%\textbf{Supplementary}
%Parametric plots of all the different mode partitions? hmmm

\textbf{Acknowledgements}

{\footnotesize This research was conducted by the Australian Research
Council Centre of Excellence for Quantum Computation and Communication
Technology (project number CE110001029) and has been supported by
the Australian Research Council DECRA and Discovery Project Grants
schemes. SA is grateful for funding from the Australia-Asia Prime
Minister`s Award. RYT thanks Swinburne University for a Research SUPRA
Award, and QHG thank National Natural Science Foundation of China
under Grant No. 11121091 and 11274025}.\\

\end{document}